\documentclass[12pt]{article}
\usepackage{graphicx}
\begin{document} 
\baselineskip 18pt

\bigskip
\centerline{\bf \large A computer simulation of language families}

\bigskip
\noindent
Paulo Murilo Castro de Oliveira$^{1,2}$, Dietrich Stauffer$^{1,3}$,
S{\o}ren Wichmann$^4$, Suzana Moss de Oliveira $^{1,2}$

\bigskip
\noindent
$^1$ Laboratoire PMMH, \'Ecole Sup\'erieure de Physique et de Chimie
Industrielles, 10 rue Vauquelin, F-75231 Paris, France

\medskip                
\noindent
$^2$ Visiting from Instituto de F\'{\i}sica, Universidade
Federal Fluminense; Av. Litor\^{a}nea s/n, Boa Viagem,
Niter\'{o}i 24210-340, RJ, Brazil

\medskip
\noindent
$^3$ Visiting from Institute for Theoretical Physics, Cologne University,
D-50923 K\"oln, Euroland

\medskip
\noindent
$^4$ Department of Linguistics, Max Planck Institute for Evolutionary
Anthropology, Deutscher Platz 6, D-04103 Leipzig, Germany \& Faculty
of Archaeology, PO Box 9515, 2300 RA Leiden, The Netherlands.


\bigskip
{\bf Abstract}

{\small 

This paper presents Monte Carlo simulations of language populations and the 
development of language families, showing how a simple model can lead to 
distributions similar to the ones observed empirically by Wichmann (2005) and others. The model combines 
features of two models used in earlier work for the simulation of 
competition among languages: the ``Viviane'' model for the migration of people 
and propagation of languages and the ``Schulze'' model, which uses bit-strings 
as a way of characterising structural features of languages.}

\section{Introduction}

In an earlier issue of this journal Wichmann (2005) showed how the sizes of 
languages families, measured in terms of the number of languages of which they 
are comprised, conform to a so-called ``power-law'' or ``Pareto distribution'', 
a special instance of which is better known to linguists as ``Zipf's law''. 
Such distributions are frequently found in both the physical and social 
universes. It was also observed, however, that the sizes of languages have a 
different kind of distribution. Wichmann called for computer simulations that 
might help us in understanding how such distributions can come about. The present paper, which 
represents the culmination of much recent work on the quantitative modelling of 
language distributions, addresses this concern. It presents simulation models 
which may help us to investigate past events leading to the current global 
language situation and which may potentially serve to simulate the future of 
global linguistic diversity.

At the time of Wichmann's writing, work on the computer simulations of the 
interaction among languages had actually already started to take flight among 
scholars in physics departments following in the footsteps of Abrams and 
Strogatz (2003). Schulze et al (2008) provide a recent review of this work (cf. 
also Wichmann et al. 2007 for a generous list of references). Moreover, a few 
years earlier, physicist Damian Zanette and biologist William Sutherland had 
respectively plotted language family sizes and language populations (Zanette 
2001, Sutherland 2003). While most simulations have been concerned with speaker 
populations, some have concentrated on modelling taxonomic structures similar 
to language families (Wang and Minett 2005, Wichmann et al. 2007, Schulze et 
al. 2008, Tuncay 2007). In spite of progress, none of the agent-based 
simulations have simultaneously captured both the current distribution of 
language sizes in terms of speaker populations (henceforth ``language sizes'') 
and the distribution of language family sizes in terms of the number of 
languages in families (henceforth ``language family sizes''). This is achieved 
in the present paper, which uses simulations of languages with internal 
structure (represented as bit-strings), and where a taxonomy of languages is 
developed through a branching mechanism starting from a single ancestor. 
The population dynamics model that we will use is based on de Oliveira et al. (2007), which has been shown to provide a good match to empirically observed distributions of numbers of speakers across the languages of the world. In this paper, an additional level of structure is added to the model, that of language families, providing a way to model empirical data about sizes of language families.

The properties of evolutionary systems can be divided into two different kinds: 
those which depend on the particular historical contingencies that have 
occurred during the evolution, and those which depend only on the general rules 
of dynamics determining how new elements of the system inherit their properties 
from other already existing elements. Such inheritance necessarily has a 
stochastic character, as is exemplified by the random genetic mutations that 
take place between parents and their offspring and which follow well-defined 
probability rules. The sequence of events can be described by a bifurcating 
historical tree, each branch corresponding to some event which has occurred in 
reality. If it were possible to return back to some remote past and to 
construct an historical evolution all over again from that point, then one 
would see a different tree evolving, even if the same rules of dynamics were
applied. Some characteristics of the new tree would differ from the real tree 
representing what has occurred in reality. Some other characteristics, however, 
are the same because both the real and the imaginary tree followed the same 
dynamic, stochastic inheritance rule. These universal characteristics relate to 
the general topology of the tree, not to whether a particular branch appears or 
not. The aim of computer models like ours is to identify and reproduce 
universal, history-independent features, simulating an artificial dynamic 
evolution. The method consists in proposing a set of stochastic inheritance 
rules, and then verifying which characteristics coincide with reality. From the 
result, one can predict some future properties which will occur independently 
of unpredictable contingencies. On the other hand, these models are not 
supposed to give any clue about details such as the particular internal 
structure of some language or language family.

\section{Family definition}

World geography is simulated by operating with a large square lattice on which 
populations can grow and migrate. We then simulate the development of 
linguistic taxa as follows (cf. the appendix for more detail). Initially, only the central point of the lattice is 
occupied by one group of people speaking one original language. This language 
(and subsequent ones) is modelled as a string of bits which can take the values 
0 or 1. These are imagined to correspond to different prominent typological 
features. The population grows and spreads over the whole lattice, with 
languages diffusing as the populations diffuse. When a new site becomes 
occupied there is a certain probability that a change occurs in one of the bits 
of the language of the population occupying the new site. If such a change 
occurs (and if the resulting bit-string is not identical with one already 
occurring elsewhere), the resulting language is defined as being a new language 
different from but descending from the language that underwent the change. 
Furthermore, with probability 1/2 this new language is defined as the starting 
point of a new language family, with all its later descendants belonging to 
this one family. If no new family is created by the new language, then all its 
later offspring again have the chance to found with probability 1/2 a new 
family, whenever another new language is created. The family founding events correspond 
to the perceived continuities in the phylogenetic landscape of the world's languages.

The definition entails three suppositions: (1) language was only created once and thus 
all languages descend from a common proto-World language; (2) linguistic diversity arises from 
changes that are stochastic in nature; (3) there are three major taxonomic levels: proto-World, 
the family level, and the language level. Assumption (1) cannot presently be proven, but is
a reasonable one, and additionally obeys Occam's razor. If assumption (2), seen as an assumption 
about the majority of linguistic changes, did not hold linguists would be able to predict 
how and when languages change, which they clearly cannot. There is also no principled way
of explaining why a certain language, such a proto-Indo-European, has ``reproductive success'' and
is subsequently recognised as a founder language by linguists some thousands of years later.
Our assumption that language changes are stochastic carries over to the process by which a founder
language is selected, which is also stochastic.
Assumption (3) is obviously reductionistic since any number of taxonomic levels could be added 
below the family level, but here we single out families and languages because these are the levels we want to investigate. Having definitions
for lower taxonomic levels (corresponding, say, to the genera of Dryer 2005, or to dialects) would 
not necessitate a different family definition, and would therefore not change the results.

A different set-up of the simulation, starting from a random point rather than 
the centre, gives similar results. One might also consider a landscape with uninhabitable
areas with mountains or oceans. Building in such features simply corresponds to
a reduction of the lattice space, which in turn corresponds to stopping the simulation
before all lattice sites are occupied. When testing effects of this we found no
differences in the results. Moreover, previous simulations of mountain ridges in the Viviane model (Schulze and Stauffer 2006) showed surprisingly little influence of the language geography. Indeed, all sorts of parameters could be added. In the somewhat different Schulze model features such as extinction of languages, migration of people, diffusion of linguistic features, the influence of geographical barriers, conquests, language shift, and bilingualism were tested (see Schulze et al. 2008 for a review). This model, however, never gave as good an agreement as figure 1 for the language size distribution. This suggests that it is the differences between the core features of our present model and the Schulze model which are important, not various aggregated parameters. 

A different definition of how a language family is created 
would be to randomly select family founders among all 
languages. Another is to consider as founders all languages of the second 
generation, counted from the ``mother tongue'' (generation zero). Another yet 
is to take random languages of the fourth generation as founders. These alternative definitions were also tested, with inferior results compared 
to the power-law exponent measured by Wichmann. Not only do these definitions not work as well,
they are also less realistic since they do not involve language change as a prerequisite
for genealogical differentiation. 
In our preferred definition a 
historical taxonomic hierarchy arises, and the resulting system of languages 
carries a long-term memory, as follows. The ``mother tongue'' is a family 
founder with certainty. Its direct descendants form the first generation, and 
each one with a 1/2 probability becomes a new family founder. Each language of 
the second generation has on average a corresponding probability 1/4, the third 
generation 1/8, etc. Therefore, the chance a new language has to become a 
family founder depends on which other languages have already founded other 
families in the past, since the very beginning. This kind of long-term memory 
is a key ingredient of various evolutionary systems having universal properties such as power-laws whose exponents 
are independent of particular contingencies occurred during the evolution, i.e., power-laws similar to that of languages family sizes.

\section{Results}

\begin{figure}[!hbt]
\begin{center}
\includegraphics[angle=-90,scale=0.73]{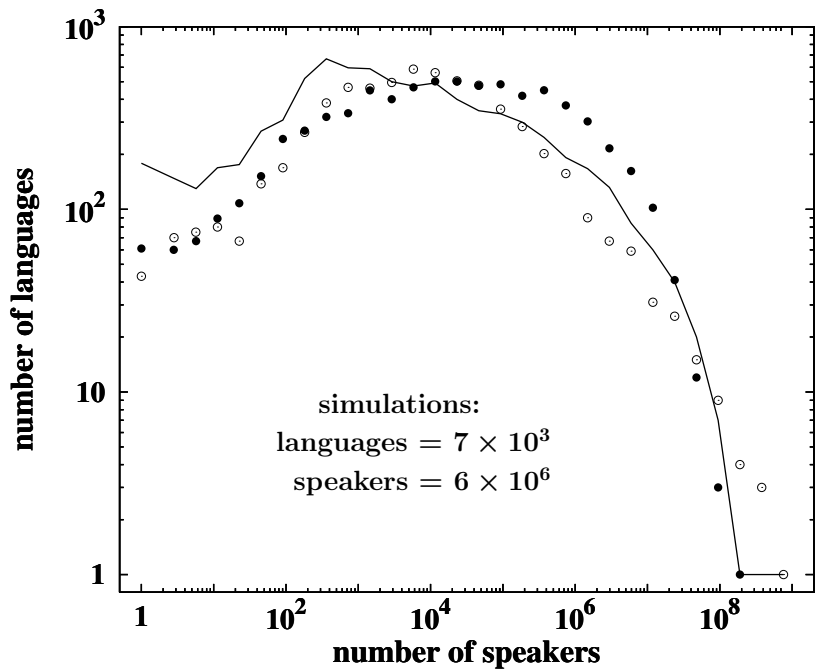}
\end{center}
\caption{Empirical size distribution of the $\sim 10^4$ present human 
languages, Grimes (2000) (open circles). The full circles show one simulation 
of our model, with parameters $L = 20,000, \; b = 13, \; M = 64, \; F_{\max} = 
256, \; \alpha = 0.07$ (see appendix). The full line corresponds to another simulation with parameters 
$L = 11,000, \; b = 16, \; M = 300, \; F_{\max} = 600, \; \alpha = 0.18$.
}
\end{figure}

The distribution of languages as a function of the number of speakers is known 
(Grimes 2000, Sutherland 2003) to be roughly log-normal, with an enhanced 
number of languages for very small sizes. Figure 1 compares reality with new 
simulations of the Viviane model (de Oliveira et al. 2006), as modified in de 
Oliveira et al. (2007), and as explained again in the appendix.

Different parameters give different curves, of which two are shown in figure 1, 
but the curves always have the same overall lognormal shape with enhancement at 
small language sizes. That is, by changing the parameters one can fine-tune 
both the height as well as the width of the curve. However, the parabolic shape 
with deviations on the left side always appears for completely different sets 
of parameters. The points on the left side represent languages spoken by very 
few people; the last point to the right represents the number of people 
speaking the largest language; and the height of the curve is related to the 
total number of languages (the integral). Within the model it is possible, for 
instance, to create a curve where the largest language is spoken by not one 
billion people but instead one million. One could also tune it to show, say, 
one thousand rather than seven thousand languages. Such adjustments, which 
might be imagined to take us back to some early stage in the evolution of 
linguistic diversity, do not change the shape of the curve, which is still 
log-normal with deviations for small languages. Thus, the overall shape of 
figure 1 is universal although its precise height or width depends on the 
numbers of speakers and languages. Different runs of simulations using one and 
the same set of parameters were also made. Deviations between different runs 
were mostly of the order of the symbol size.

Once parameters 
were fitted to produce the results for language sizes shown in figure 1 they 
were not adjusted further in order to capture the family size distributions. 
The latter followed directly from the same settings which produce the full 
circles in figure 1. 

\begin{figure}[!hbt]
\begin{center}
\includegraphics[angle=-90,scale=0.5]{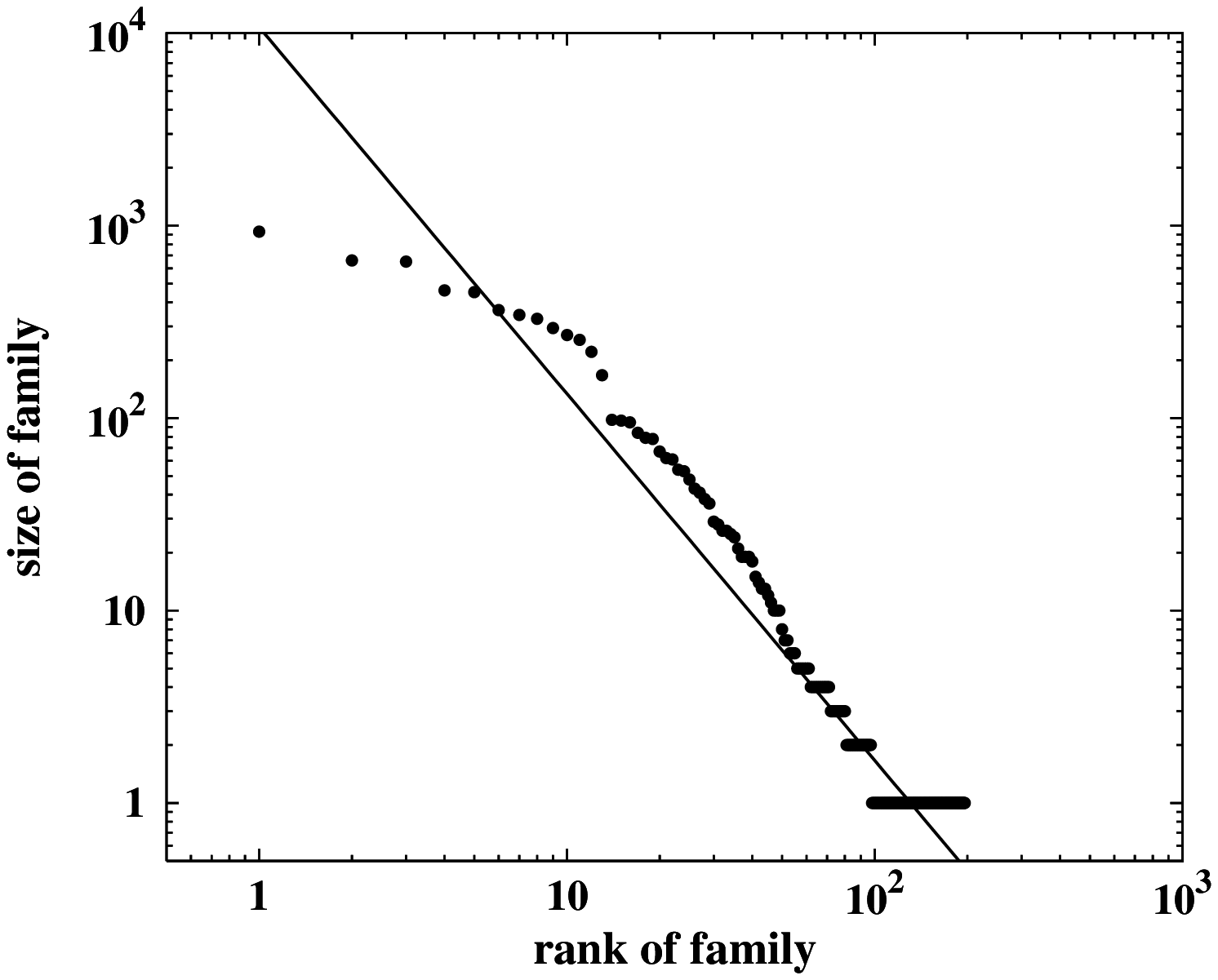}
\vskip 0.5cm\hskip-20pt
\includegraphics[angle=-90,scale=0.53]{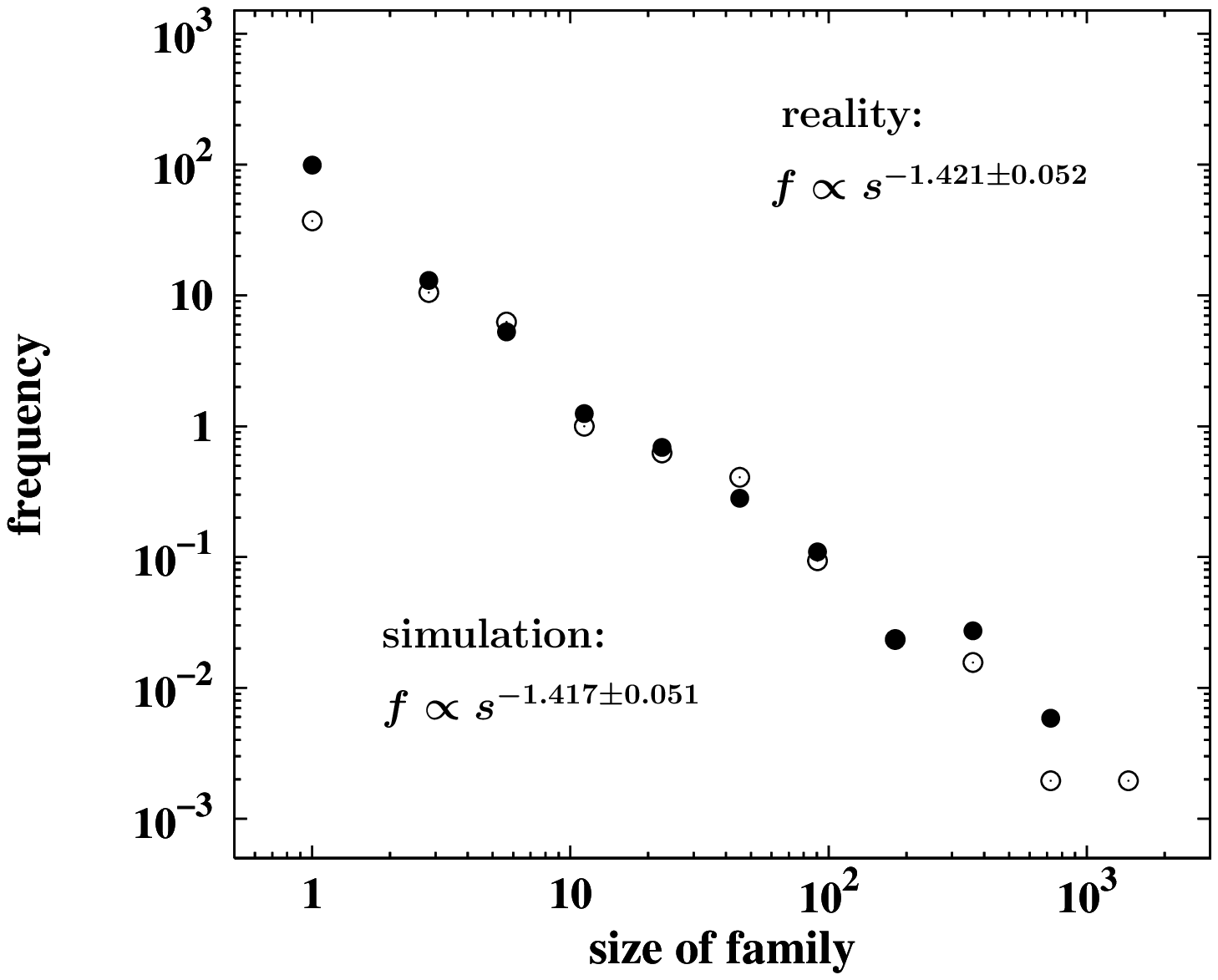}
\end{center}
\caption{Number of languages in a family. The straight line is not a fit on 
these data but the fit of Wichmann (2005) on his rank plot taken from
real languages Grimes (2000). In the lower plot, full circles are simulated
data points and open circles empirical data points.
}
\end{figure}

\begin{figure}[!hbt] \begin{center} 
\includegraphics[angle=-90,scale=0.33]{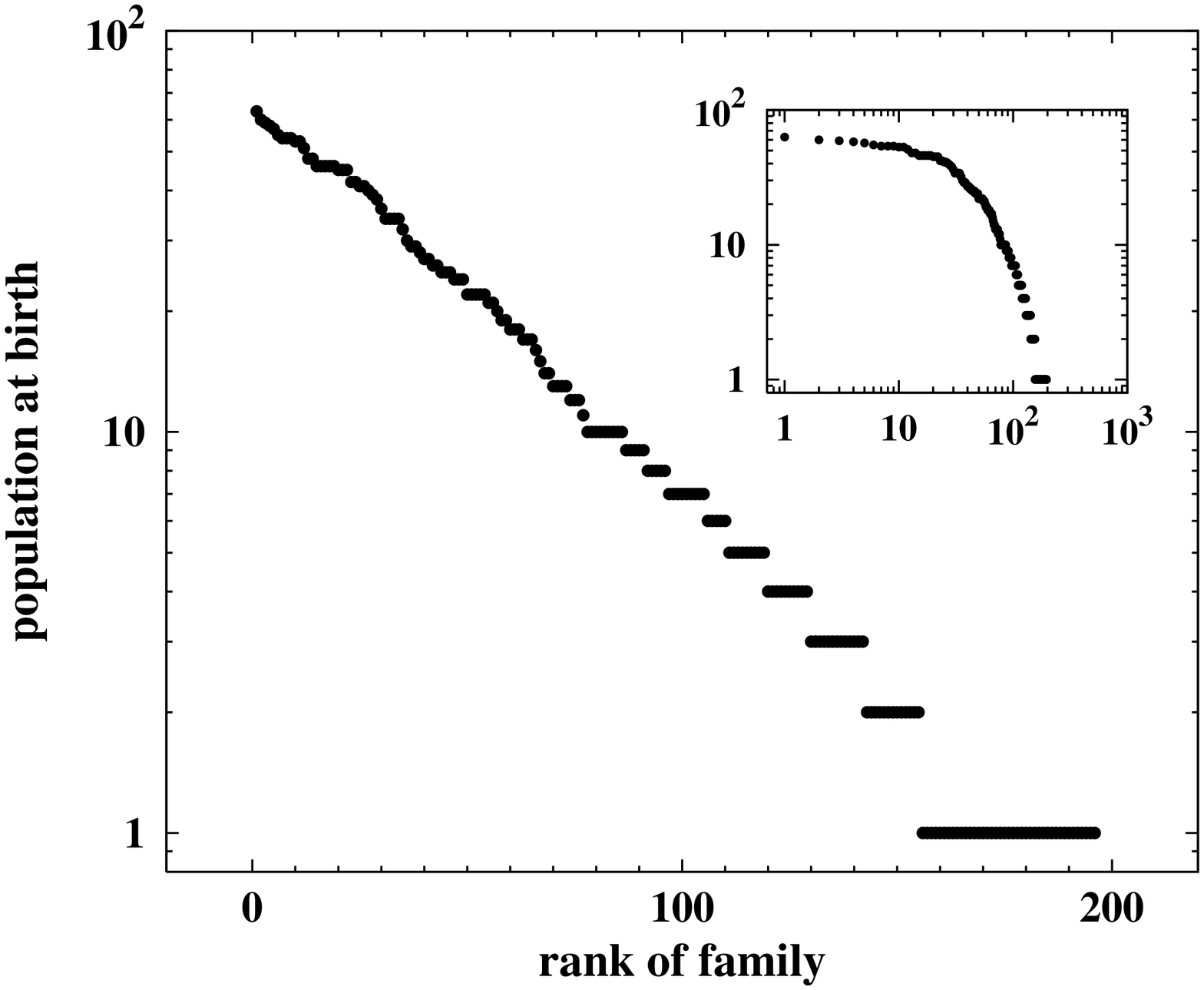} 
\includegraphics[angle=-90,scale=0.33]{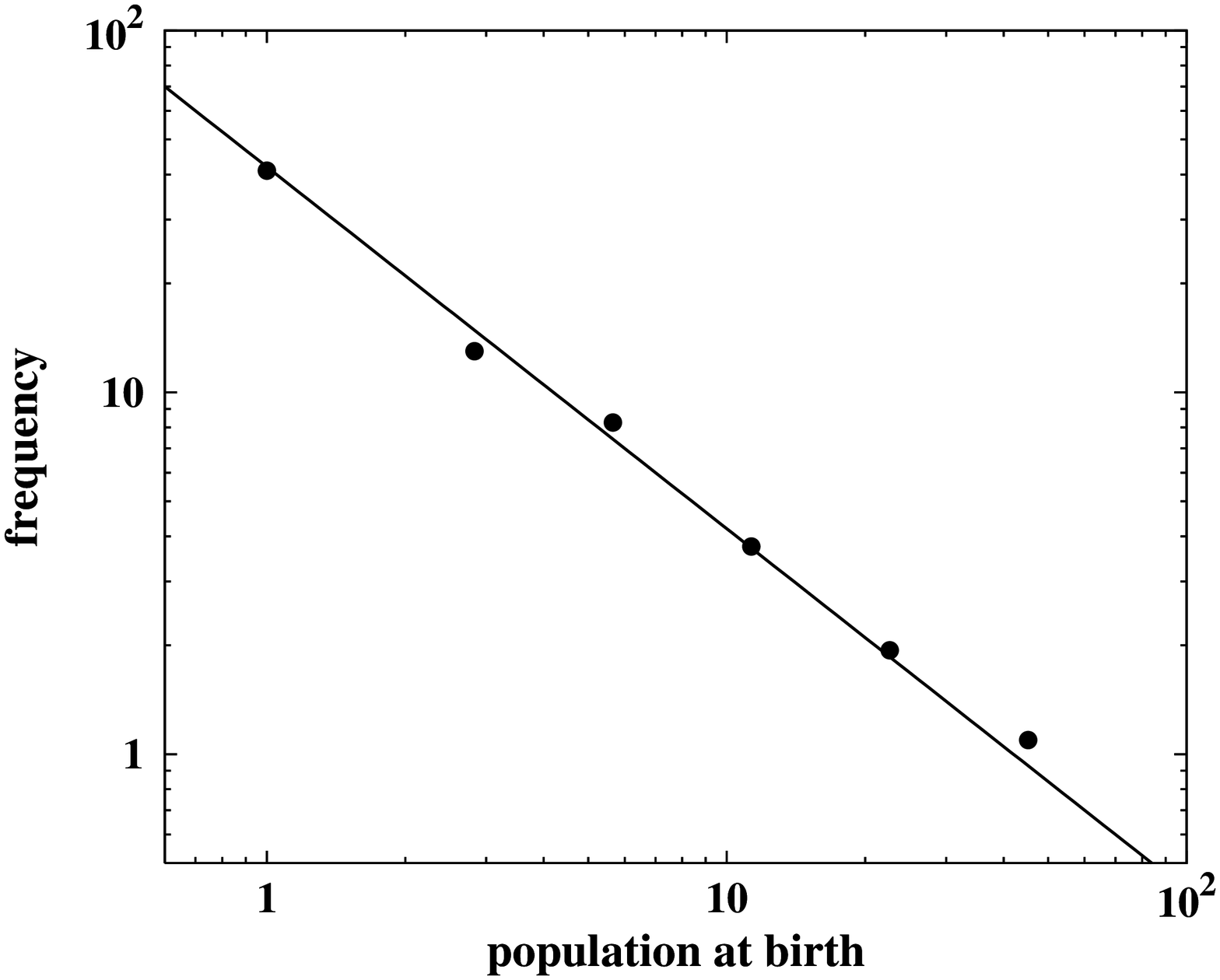}
\end{center}

\caption{Initial population of the founder of a family. 
Ranking in the upper plot is by population size. Different from the 
log-log plot, now the ranking was displayed with linear horizontal scale, for 
which the straight behaviour shown in the upper plot indicates an exponential 
decay. The inset here (same for figures 4-5) shows the corresponding log-log curved plot. 
Accordingly, the straight line on the frequency plot (below) gives $\tau = 1$.}

\end{figure}

\begin{figure}[!hbt]
\begin{center}
\includegraphics[angle=-90,scale=0.33]{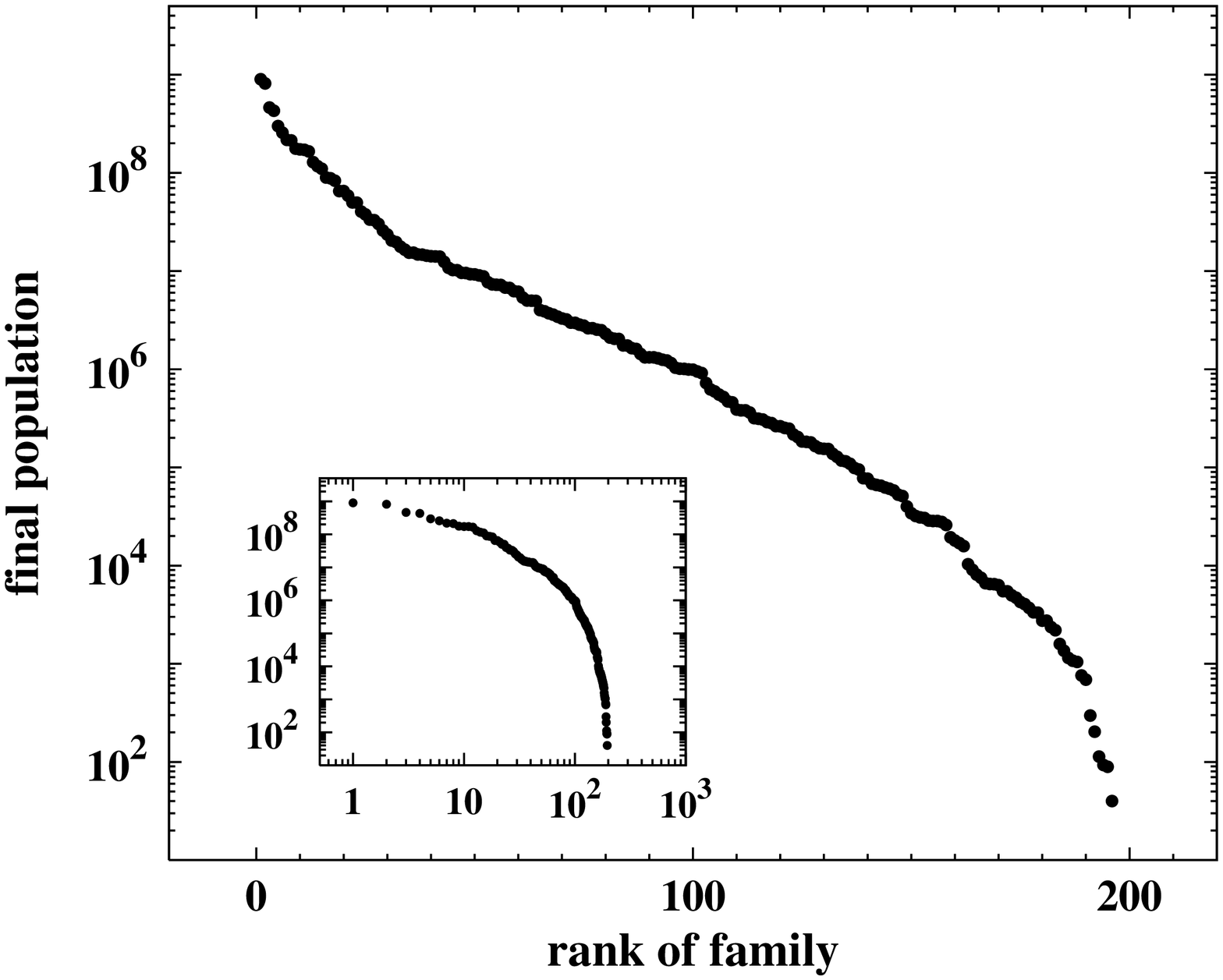}
\includegraphics[angle=-90,scale=0.33]{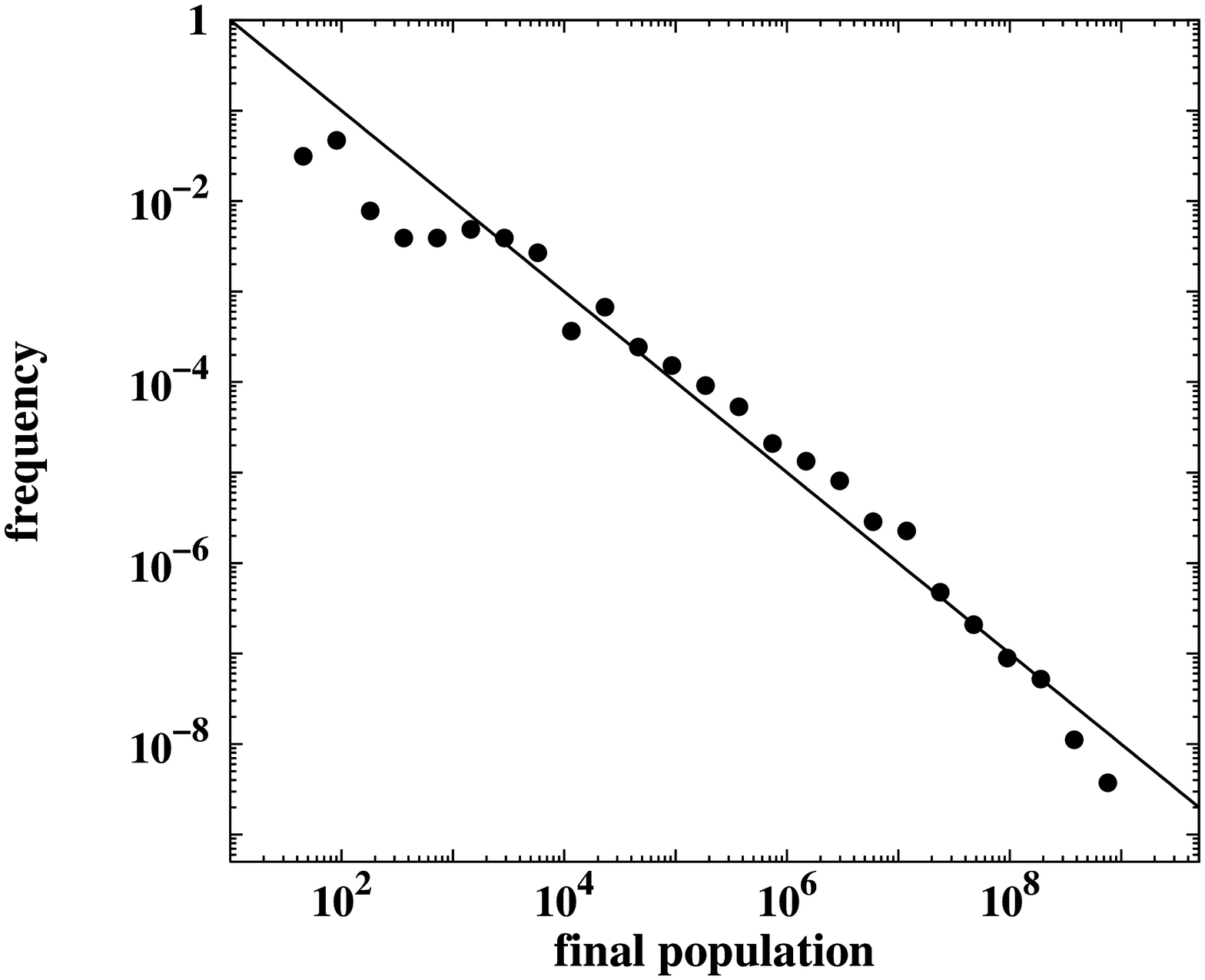}
\end{center}
\caption{Number of speakers in a family (ranking is by final population size).}
\end{figure}

\begin{figure}[!hbt]
\begin{center}
\includegraphics[angle=-90,scale=0.33]{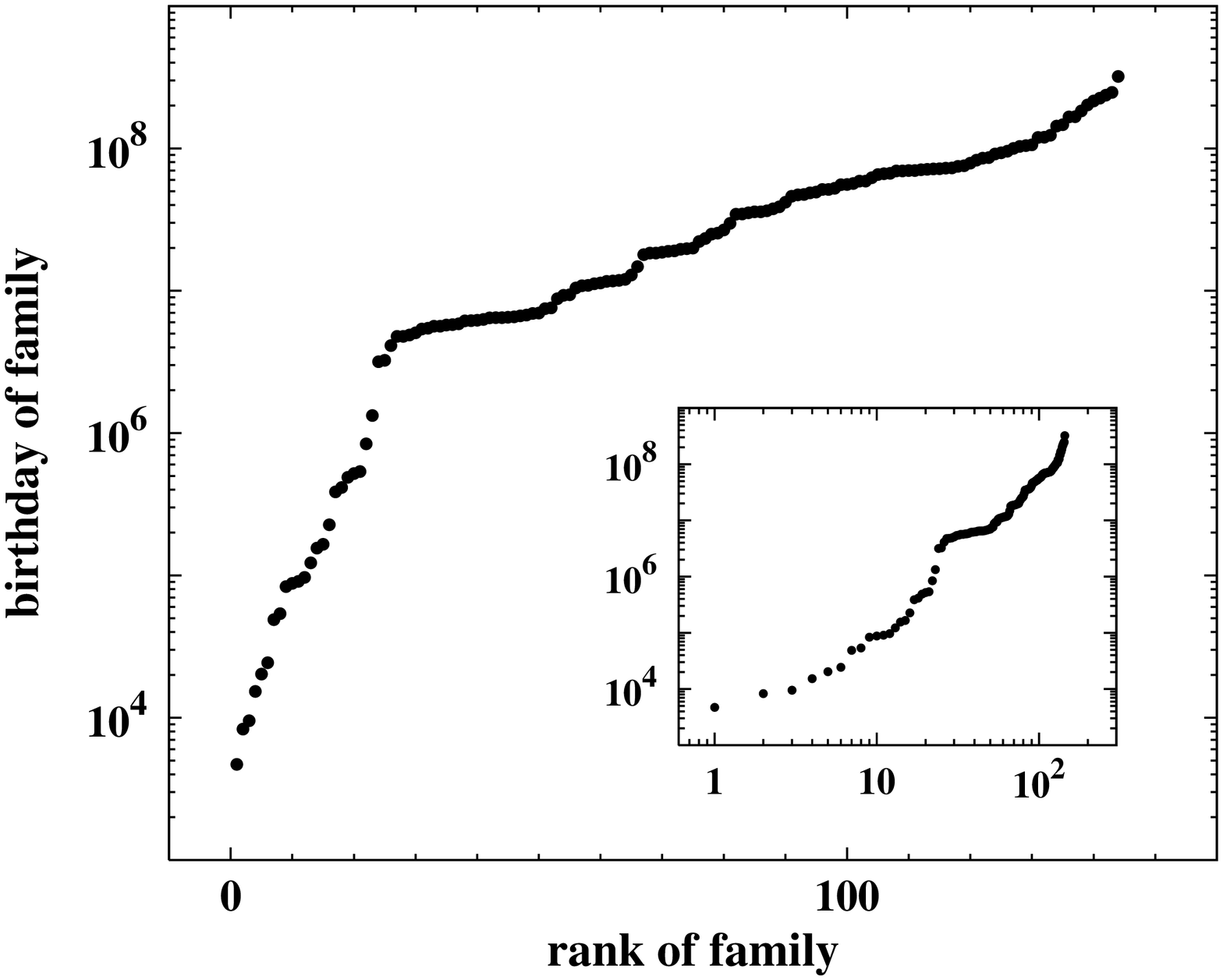}
\includegraphics[angle=-90,scale=0.33]{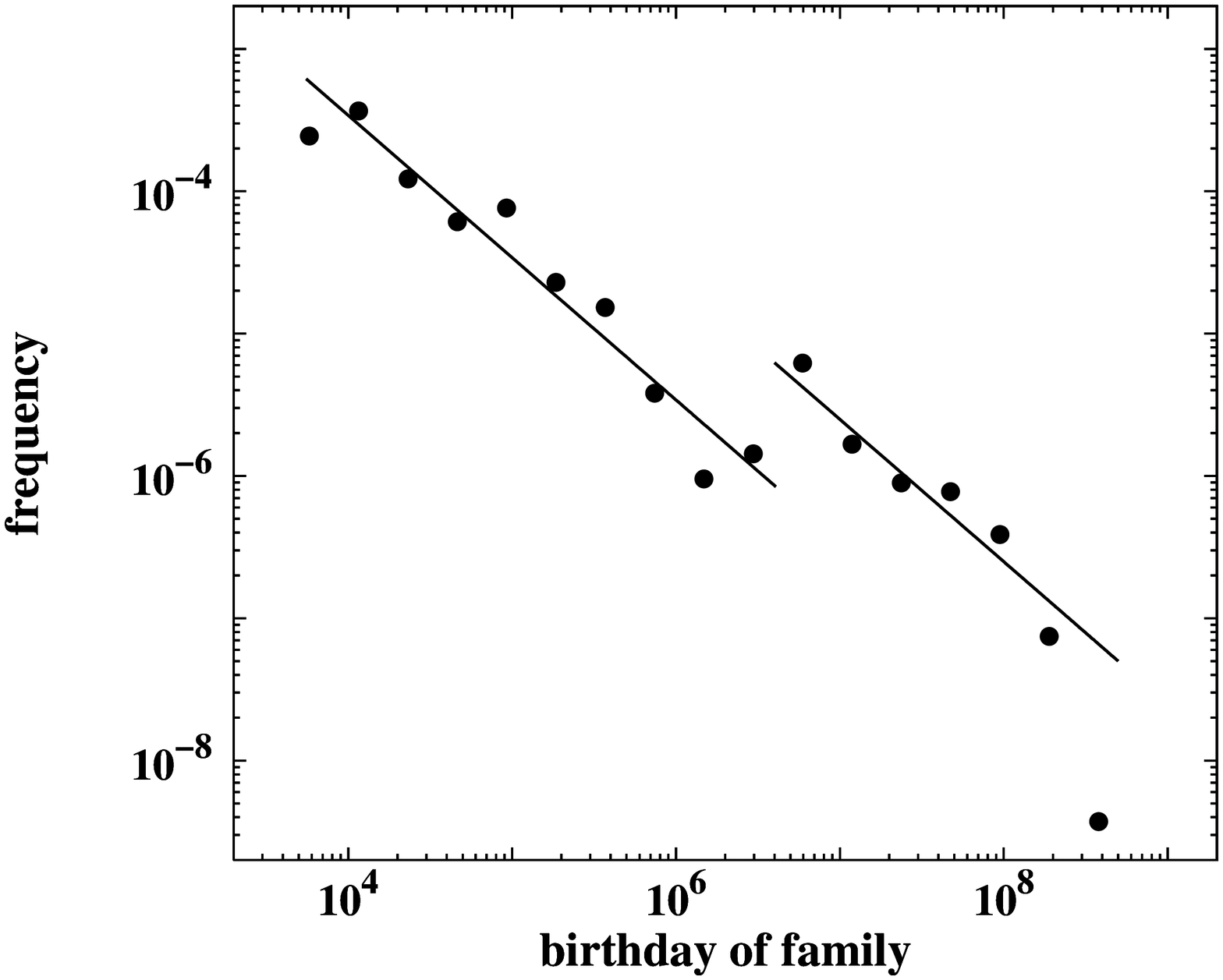}
\end{center}
\caption{Birthday of a family (ranking is by birthday).}
\end{figure}

\begin{figure}[!hbt]
\begin{center}
\includegraphics[angle=-90,scale=0.33]{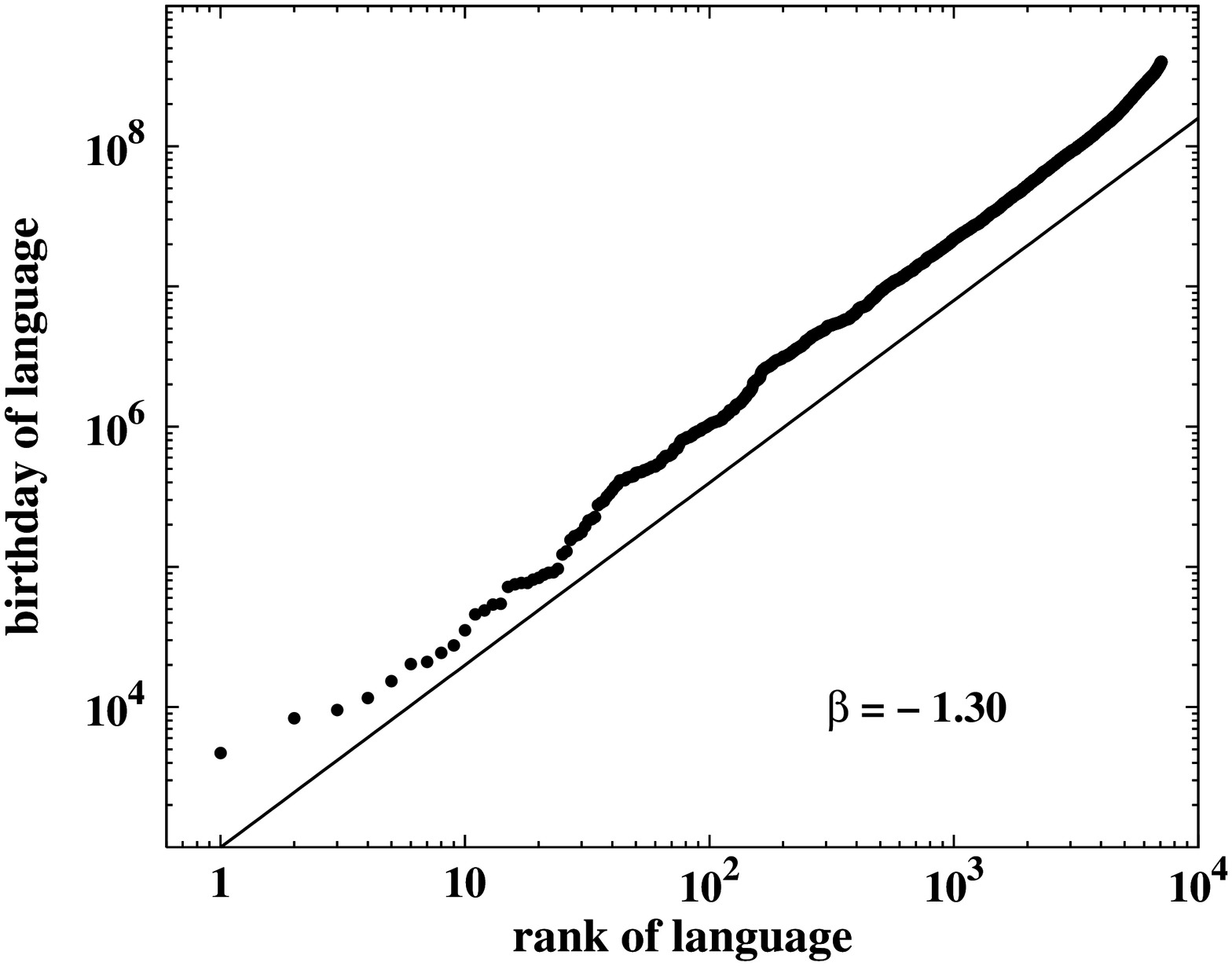}
\includegraphics[angle=-90,scale=0.33]{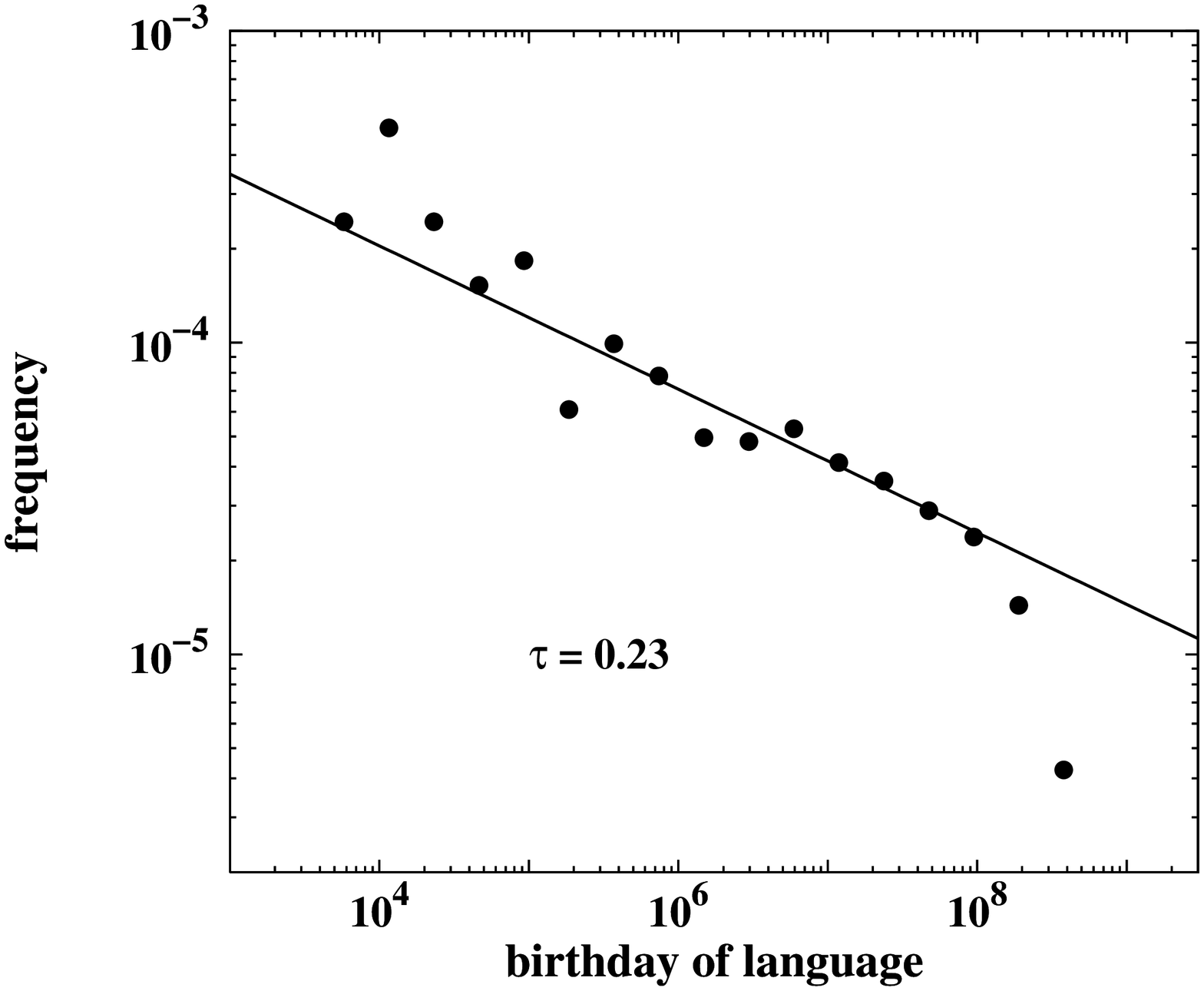}
\end{center}
\caption{Birthday of a language (not family) (ranking by birthday).}
\end{figure}

\begin{figure}[hbt]
\begin{center}
\includegraphics[angle=-90,scale=0.5]{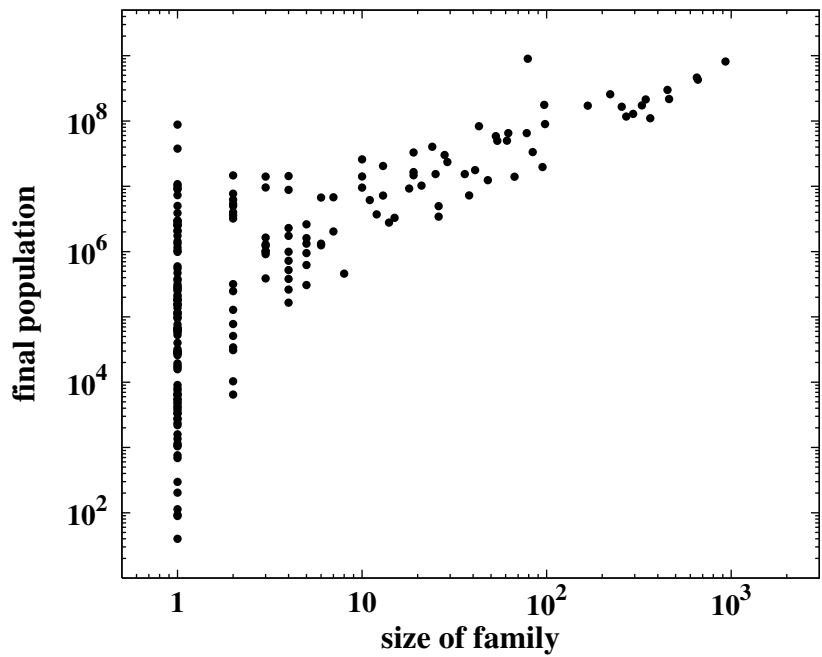}
\end{center}
\caption{Strong correlation between family population and family size.
Each point corresponds to a family.
Neither averaging nor binning is used in the scatter plots of figures 7 to 9.
}
\end{figure}

\begin{figure}[hbt]
\begin{center}
\includegraphics[angle=-90,scale=0.5]{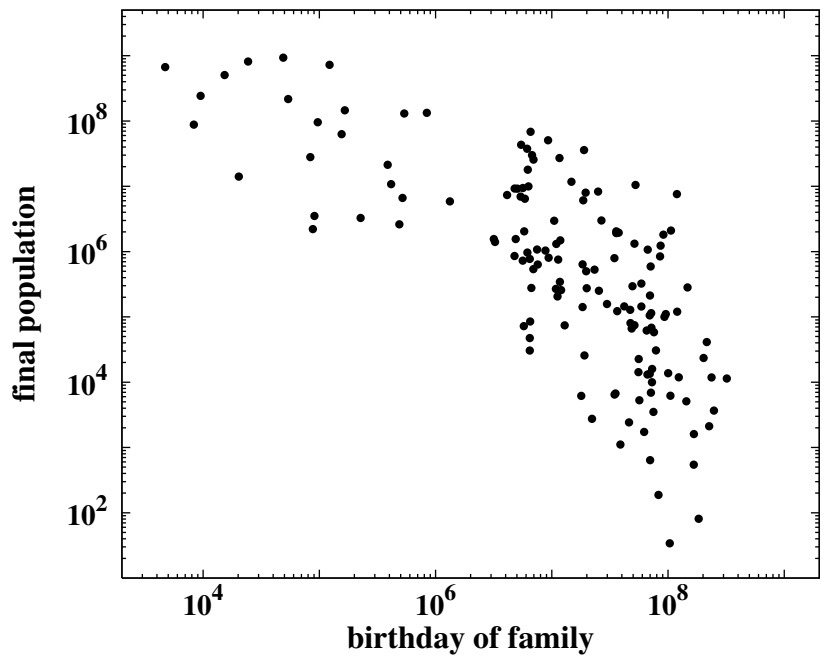}
\end{center}
\caption{Strong correlation between family birthday and family population.
}
\end{figure}

\begin{figure}[hbt]
\begin{center}
\includegraphics[angle=-90,scale=0.5]{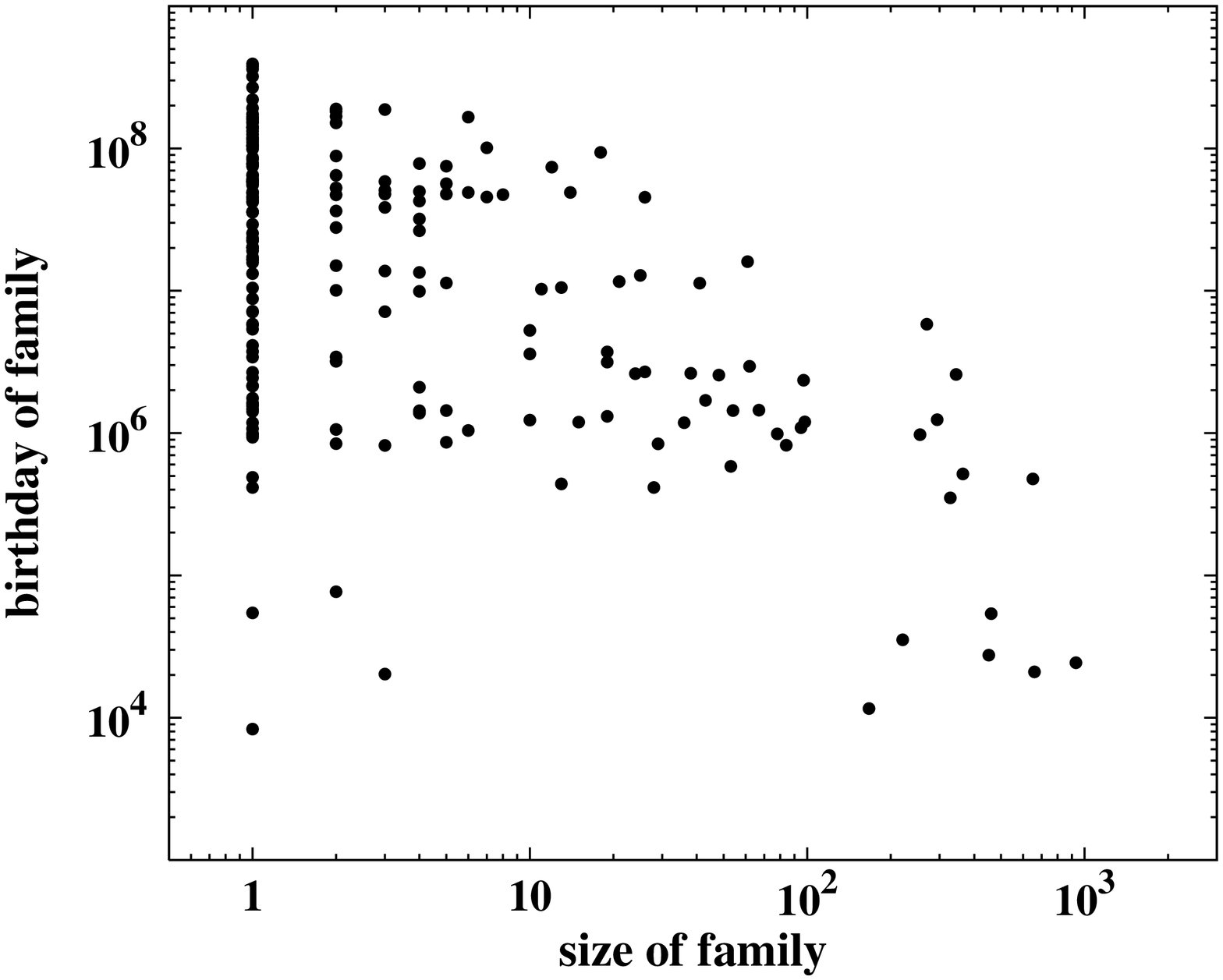}
\end{center}
\caption{Weak correlation between family size and family birthday.
}
\end{figure}

The plots in figures 2-6 always consist of two parts: a rank plot on top and a 
histogram below it. For example, for the size (= number of languages in a 
language family) the rank plot shows on its left end the largest family, 
followed by the second-largest family, then the third-largest family, etc. The 
histogram below shows on its left end the number of families containing only 
one language (``isolates''), followed by those containing two, three, and more 
languages. To avoid overcrowding in the plots, we binned sizes together by 
factors of two, that means sizes 2 and 3 give one point, all sizes from 4 to 7 
give the next point, all sizes from 8 to 15 the next, etc; the resulting sum is 
divided by the length 2, 4, 8, ... of the binning interval and gives the 
frequency. This division is not made in figure 1, which gives the summed 
numbers. If the rank plot is described by a power-law $s \propto r^{-\beta}$ 
(where the symbol $\propto$ represents proportionality), then the corresponding 
frequency plot is also described by another power-law $f \propto s^{-\tau}$, 
where $\beta = 1/(\tau-1)$. In the particular case of $\tau = 1$ the 
corresponding rank plot is no longer described by a power-law, but by an 
exponential function $s \propto {\rm exp}(\lambda r)$.

Figure 2 gives the number of languages in each family. Figure 3 shows 
the population of each language at the site where it gave rise to a new 
family. Figure 4 gives the number of speakers in each family. This turns 
out to be proportional to the number of lattice sites occupied by the 
speakers of each family (not shown). Finally, figure 5 shows the birthday (number of iterations since the 
start of the simulation) of each family. In all cases the histogram 
roughly follows a power-law (straight line in our log-log plots), and 
figure 2, our most important plot, shows that also the rank plot follows 
a power-law compatible with Wichmann's exponent 1.905. The histograms 
are more sensitive tests of the power-laws than the rank plot, for both 
reality and simulations.

These power-laws are not valid over the whole range (Arnold and Bauer 2006), 
either in our simulations or in reality: No family can contain half a language, 
or more than the total $10^4$ languages. But the exponents in the central part 
are not only a convenient way to summarise results in one number; they also 
seem to have some universality in the sense that the same exponent tends to 
occur independently of many details of the simulations. Indeed, when we changed parameters (including the probability 1/2 of 
Section 2) the details of our results changed but the central exponents did not 
change significantly. 

Only the definition of families had drastic effects on the outcome. As 
mentioned above, we tried other possible definitions. However, only the 
hierarchical definition presented in Section 2 gives the proper exponents 
compared with reality, figure 2. The variation in results from different 
definitions suggests that continuous branching is the most realistic 
description of the evolution that has led to the present phylogenetic 
diversity.

Figure 5 presents a curious behaviour. Instead of a single straight line, the 
ranking plot consists of two, which correspond to $s \propto {\rm 
exp}(\lambda_1 r)$ for the first oldest families and ${\rm exp}(\lambda_2 r)$ 
for the more recent ones, with $\lambda_1 > \lambda_2$. This transition from 
one regime to the other defines a typical time scale when the successive 
creation of new families changes rhythm such that the quantity of new families 
formed per time unit increases. It also appears for different sets of 
parameters and/or random numbers we tested. In the frequency plot, the 
signature of this transition is the presence of two parallel straight lines, 
both corresponding to $\tau = 1$. The explanation for the knee in the upper 
plot of figure 5 relates to the fact that the simulations start from a single 
ancestor. The production of new founders is relatively slow in the beginning 
when there are only few branches on the tree, but when the tree gets 
sufficiently complex the dynamics changes and founders are produced at shorter 
intervals. To test whether something similar to the knee of figure 5 occurs in 
reality we plotted the data for cognate percentages for most of the world's 
languages families which were collected by Holman (2004) from a variety of 
sources. If the assumptions of glottochronology are correct these cognate 
percentages should translate into ages. A curve with a shape similar to that of 
figure 5 results, also having a ``knee'', even if only three families are found 
in the lower part of the ``leg'': Afro-Asiatic (6\% cognates), Eastern Sudanic 
(9\% cognates), and Chibchan (11\% cognates). Thus the tendency is not so 
pronounced. The explanation for this ``empirical knee'' may be the same as for 
the behaviour of the simulations, supporting the idea that all language families 
derive from a common ancestor. It is equally possible, however, that the 
explanation relates to the fact that it gets more difficult to establish what 
is and what is not a cognate as the time depth increases; the deviant behaviour 
for a few old families, then, could be due to fluctuations in knowledge.

The rhythm of successive appearance of new languages (not families), as shown 
in figure 6, does not exhibit the kind of transition between two regimes that 
we saw in relation to families. Instead, both the ranking and the frequency 
plot seem to be described by power-laws.

We also looked at correlations between the various results. Area and population 
are proportional to each other apart from statistical fluctuations, as 
expected. It is also plausible that the final population increases with the 
size of the family (figure 7), and decreases with the birthday of the family 
(figure 8), both in a nonlinear way. Figure 9 shows only a weak correlation 
between birthday (age) and family size. This is compatible with reality, where 
the size of a language family is not necessarily an indicator of its age.

Using a slightly different program, we found that the average number of 
generations from a final language back to the one original language increases 
about logarithmically for large lattice sizes but more weakly for small 
lattices. In all of the above versions the language at one site never changes 
after the site becomes inhabited. Instead, we also included a later diffusion 
of language features to and from already occupied neighbour sites, for all or 
for only selected bit positions. Then for strong diffusion we found a strong 
reduction of the number of languages, without a drastic change in the family 
size histogram.

\section{Outlook}

Our simulations gave a surprisingly good agreement with reality for the 
rank plot of family sizes, cf. figure 2a. The number of languages as a 
function of occupied area was already found in earlier work (de Oliveira 
et al. 2006) to agree with reality (Nettle 1998). Since one and the same 
model can produce both the current language size and family size 
distributions these two distributions are not likely to be somehow out 
of tune due to the current rapid extinction of many languages---a 
possibility very tentatively raised by Wichmann (2005: 128).

Given that the model is sufficiently fine-tuned to capture the quantitative 
distributions just mentioned it may be considered an adequate starting-point 
for addressing other problem areas that invite simulations. Unlike some other 
models that operate with languages without internal structure the combined 
Schulze-Viviane model characterises languages in terms of bit-strings. For 
instance, this makes it possible to use the model for testing how well 
different phylogenetic algorithms can adequately recuperate taxonomic relations 
among languages from the distributions of their typological features (cf. 
Wichmann and Saunders 2007). Other issues of language change may be addressed, 
such as the development and distribution of creoles, large-scale diffusion of 
linguistic features, change rates of typological profiles, prehistoric 
bottle-neck effects, and last, but not least, the future of global linguistic diversity. We see the development of a simulation model which is 
both simple and versatile as the most important outcome of the present 
contribution.

In this paper we have simulated sizes of language families and populations. 
Whether one language or language family grows or shrinks depends on many 
historical events which we have not taken into account, such as wars, famines, 
etc. While such individual events are not predictable, we know from other 
social and physical phenomena that after a long history of interaction among 
many components of a system overall statistical properties emerge which are 
independent of specific events of the process. Thus, it does make sense to 
simulate on a computer how many languages belong to the largest family, how 
many to the second-largest family, etc, without specifying which family is the 
largest, or what rank a given family, such as Indo-European or other, has. The 
evolution (of living beings, languages, etc.) depends on the particular 
sequence of historical events, and contingencies having occurred at some past 
influence the future. However, for statistics involving thousands of elements, 
the structure of an evolutionary trajectory presents some basic universal 
characteristics which are independent of the particular contingencies that have 
occurred in reality and depend only on these contingencies having occurred 
according to some prescribed probability rules common for different kinds of 
evolutionary systems.

\section{Appendix: Modified Viviane model}

The Viviane model of language competition, as modified in de Oliveira et al. 
(2007) describes the spread of human population over a previously uninhabited 
continent. Each site $j$ of a large $L \times L$ lattice can carry a population 
$c_j$, chosen randomly between 1 and a maximum $M$, with a probability 
inversely proportional to $c$ for large $c$, more precisely $c = 
\exp[r*\ln(M)]$, where $r$ is a random number between 0 and 1. On each site 
only one language is spoken, characterised by a string of $b$ bits (0 or 1). 
Initially only the central lattice site is occupied. Then at each iteration, 
one empty neighbour $j$ of the set of unoccupied sites becomes populated by 
$c_j$ people. This newly inhabited site is selected by randomly choosing two 
empty neighbours of the set of occupied sites and by taking the one with the 
larger $c$. The new site gets the language $\ell$ of one of the occupied 
neighbours $i$, selected with a probability proportional to the fitness of this 
language. This fitness $F_\ell$ is the number of people speaking at that time 
the same language $\ell$ spoken at site $i$, bounded from above by some maximum 
fitness chosen randomly between 1 and $F_{\max}$. Once the new site $j$ is 
occupied, its language $\ell$ changes with probability $\alpha/F_\ell$, with 
some proportionality factor $\alpha$. Such a change means that one randomly 
selected bit is changed.  The simulation stops if all sites became occupied; 
the total number of languages is then the total number of different 
bit-strings.

[NOTE ADDED IN PROOF:
The assumption here is that the language change rate is inversely proportional 
to the population size. Recent work on empirical data carried out with Eric W. 
Holman suggests that this assumption is questionable. Therefore, as the present 
paper is going to press, we have made additional simulations were the rate of
language change and the occupation of a new site are independent of the number
of speakers of the languqge; these gave frequency distributions of language
and family sizes similar to Figures 1 and 2, showing that assumptions
about the relation between the population sizes and the language change 
rates are unimportant for the results of our model.]

Figure 10 provides snapshots of the gradual occupation of the 
lattice. The figure is included for illustrative purposes only, so the 
lattice contains only $20 \times 20$ sites. At 50 time steps we see the 
distribution of the initial language (open circles) and the birth of a 
second one (asterix). The sizes of the symbols correspond to the 
population at each site. At 150 time steps yet a third (black square) 
and a fourth (black circle) language have been born. At 250 time steps 
we see the further expansions of previously born languages and the 
coming about of some new ones (right and left triangles). The final 
snapshot shows the the fully occupied lattice with yet more new symbols 
for new languages, and a total of 12 languages.

\begin{figure}[hbt!]
\begin{center}
\includegraphics[angle=-90,scale=0.7]{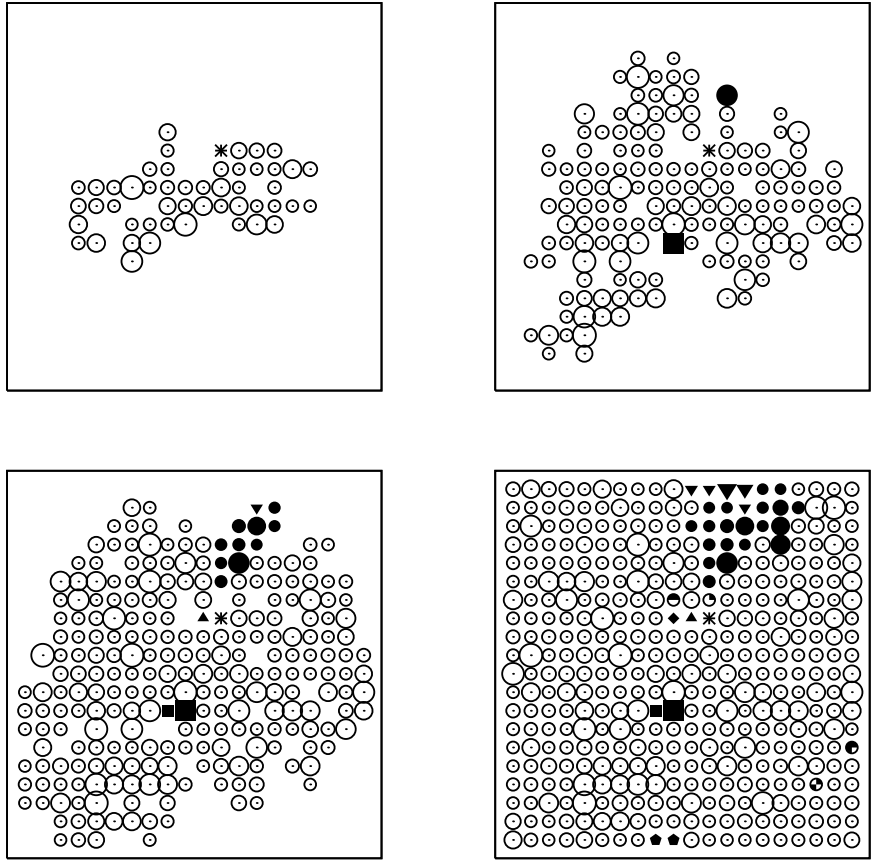}
\end{center}
\caption{Snapshots of the growth of a small lattice.
}
\end{figure}

While parameters may be varied to fine-tune the results with reality the 
parameters themselves cannot be translated into or adjusted to reality 
since they are all quite abstract. The model of the spread and 
competition among languages, on the other hand, does carry assumptions 
about how things work in reality. The preference for people to spread to 
sites with higher carrying capacities mirrors the preference for areas 
with better food resources. Further, larger languages are seen as having 
a better chance of spreading than smaller ones. These assumptions are 
hardly controversial. The fact that the probability for a language to 
change is inversely proportional to the total number of speakers of the 
language (limited by an upper bound) may be more controversial, but is 
supported by Nettle (1999) and finds some further support from both 
empirical data and simulations in Wichmann et al. (forthc.).

\bigskip
\bigskip
\noindent
{\bf \Large References}

\medskip
\noindent
Abrams, Daniel and Steven H. Strogatz. 2003. Modelling the dynamics of
language death. \textit{Nature} 424: 900.

\medskip
\noindent
Arnold, Richard and Laurie Bauer. 2006. A note regarding 
``On the power-law distribution of language family sizes.'' \textit{Journal of 
Linguistics} 42: 373-376.

\medskip \noindent de Oliveira, Viviane M., Marcelo A. F. Gomes, and Ing Ren 
Tsang. 2006. Theoretical model for the evolution of the linguistic diversity. 
\textit{Physica A} 361: 361-370; de Oliveira, Viviane M., Paulo R.A. Campos, 
Marcelo A. F. Gomes, and Ing Ren Tsang. 2006. Bounded fitness landscapes and 
the evolution of the linguistic diversity. \textit{Physica A} 368: 257-261.

\medskip
\noindent
de Oliveira, Paulo Murilo Castro, Dietrich Stauffer, F. Welington S. 
Lima, Adriano de Oliveira Sousa, Christian Schulze, and Suzana Moss de 
Oliveira. 2007. Bit-strings and other modifications of Viviane model for 
language competition. \textit{Physica A} 376: 609-616. Preprint: 
0608.0204 on arXiv.org.

\medskip
\noindent
Dryer, Matthew S. 2005. Genealogical language list. In Haspelmath, 
Martin, Matthew Dryer, David Gil, and Bernard Comrie (eds.) \textit{The 
World Atlas of Language Structures}, 584-643. Oxford: Oxford University 
Press.

\medskip
\noindent
Grimes, Barbara F. 2000, \textit{Ethnologue: languages of the world} (14th edn. 
2000). Dallas, TX: Summer Institute of Linguistics; www.sil.org.

\medskip
\noindent
Holman, Eric W. 2004. Why are language families larger in some regions 
than in others? \textit{Diachronica} 21: 57-84.

\medskip
\noindent
Nettle, Daniel. 1998. Explaining global patterns of language diversity. 
\textit{Journal of Anthropological Archaeology} 17: 354-374.

\medskip
\noindent
Nettle, Daniel. 1999. Is the rate of linguistic change constant? 
\textit{Lingua} 108: 119-136.

\medskip
\noindent
Schulze, Christian and Dietrich Stauffer. 2006. Recent developments in computer simulations of language competition. \textit{Computing in Science and Engineering} 8: 86-93.

\medskip
\noindent
Schulze, Christian, Dietrich Stauffer, and S{\o}ren Wichmann. 2008. 
Birth, survival and death of languages by Monte Carlo simulation. 
\textit{Communications in Computational Physics} 3: 271-294. Preprint: 
0704.0691 on arXiv.org.

\medskip
\noindent
Sutherland, William J. 2003. Parallel extinction risk and global distribution 
of languages and species. \textit{Nature} 423: 276-279.

\medskip
\noindent
Tuncay, \c{C}a\u{g}lar. 2007. Physics of randomness and regularities for 
cities, languages, and their lifetimes and family trees. \textit{International 
Journal of Modern Physics C} 18: 1641-1658. Preprint: 0705.1838 on arXiv.org.

\medskip
\noindent
Wang, William S.Y. and James W. Minett. 2005. Vertical and horizontal 
transmission in language evolution. \textit{Transactions of the Philological 
Society} 103: 121-146.

\medskip
\noindent
Wichmann, S{\o}ren. 2005. On the power-law distribution of language family sizes. 
\textit{Journal of Linguistics} 41: 117-131.

\medskip
\noindent
Wichmann, S{\o}ren and Arpiar Saunders. 2007. How to use typological databases 
in historical linguistic research. \textit{Diachronica} 24: 373-404.

\medskip
\noindent
Wichmann, S{\o}ren, Dietrich Stauffer, F. Welington S. Lima, and Christian 
Schulze. 2007. Modelling linguistic taxonomic dynamics. \textit{Transactions of 
the Philological Society} 105.2: 126-147

\medskip
\noindent
Wichmann, S{\o}ren, Dietrich Stauffer, Christian Schulze, Eric W. Holman. 2008. Do language change rates depend on population size? \textit{Advances in Complex Systems} 11.3: 357-369. Preprint: 0706.1842 on arXiv.org.

\medskip
\noindent
Zanette, Damian. 2001. Self-similarity in the taxonomic classification of human 
languages. \textit{Advances in Complex Systems} 4: 281-286.

\end{document}